\documentclass[a4paper,11pt]{article}
\pdfoutput=1

\usepackage[utf8]{inputenc}
\usepackage[english]{babel}
\usepackage{geometry}
\usepackage{amsmath}
\usepackage{amsfonts}
\usepackage{bbm}
\usepackage{amsthm}
\usepackage{thmtools}
\usepackage{mathtools}
\usepackage{authblk}
\usepackage[authoryear]{natbib}
\usepackage{tikz}
\usepackage{tikzscale}
\usepackage{caption}
\usepackage{subcaption}
\usepackage{pgfplots,pgfplotstable}
\usepackage{csquotes}

\declaretheorem[name={Algorithm},numberwithin=section]{myalgo}
\mathtoolsset{showonlyrefs}
\newtheorem{proposition}{Proposition}
\tikzset{font={\fontsize{8pt}{10}\selectfont}}
\title{  \bf Permutation tests for the equality of covariance operators of functional data with applications to evolutionary biology}

\author{Alessandra Cabassi}
\affil{MRC Biostatistics Unit, School of Clinical Medicine, University of Cambridge, Cambridge, UK}

\author{Davide Pigoli\footnote{Address for correspondence: Davide Pigoli,
Statistical laboratory, Department of Pure Mathematics and
Mathematical Statistics, University of Cambridge, Wilberforce Road,
Cambridge, CB3 0WB, United Kingdom. Email:
dp497@cam.ac.uk}}
\affil{Statistical Laboratory, University of Cambridge, Cambridge, UK}
\author{Piercesare Secchi}
\affil{Department of Mathematics, Politecnico di Milano, Milan, Italy}

\author{Patrick A. Carter}
\affil{School of Biological Sciences, Washington State University, Pullman, USA}
\date{}

\begin{document}

\maketitle

\begin{abstract}
In this paper, we generalize the metric-based permutation test for the equality of covariance
operators proposed by \citet{pigoli2014distances} to the case of multiple samples of functional
data. To this end, the non-parametric combination methodology of \citet{pesarin2010permutation} is
used to combine all the pairwise comparisons between samples into a global test. Different
combining functions and permutation strategies are reviewed and analyzed in detail. The resulting
test allows to make inference on the equality of the covariance operators of multiple groups and,
if there is evidence to reject the null hypothesis, to identify the pairs of groups having
different covariances.
% Multiplicity control
It is shown that, for some combining functions, step-down adjusting procedures are available to
control for the multiple testing problem in this setting.
% Simulations
The empirical power of this new test is then explored via simulations and compared with those of
existing alternative approaches in different scenarios.
% Application
Finally, the proposed methodology is applied to data from wheel running activity experiments, that
used selective breeding to study the evolution of locomotor behavior in mice.
\end{abstract}


% history:
% \received{\smonth{1} \syear{0000}}

%\tableofcontents
\section{Introduction}
In recent years, an increasing number of applications has involved data that are best described as
being functional. Examples can be found in medicine (\citealp{west2007functional}), neuroimaging
(\citealp{jiang2009smoothing}, \citealp{viviani2005functional}), biology
(\citealp{wu2010functional}, \citealp{illian2009functional}), finance
(\citealp{laukaitis2008functional}) and quality control (\citealp{colosimo2010comparison},
\citealp{torres2011detection}), to mention just a few fields.

%When applying traditional statistical methods to this kind of data, two issues arise: not only the the number of variables of interest is much larger than the number of observations (so-called large $p$ small $n$ problem), but there are also strong correlations between the covariates.
These data asked for the development of new methodologies that take into account the properties of
the functional data (see \citealp{ramsay2005functional}, \citealp{ferraty2006nonparametric} and
\citealp{horvath2012inference}). Most recently, much attention has been devoted to inferential
procedures for covariance operators of functional data. \citet{panaretos2010second} examined the
testing of equality of covariance structures from two groups of functional curves generated from
Gaussian processes and \citet{fremdt2013testing} extended their approach to the case of non
Gaussian data. Both methods make use of test statistics based on the Karhunen--Lo\'{e}ve expansions
of the covariance operators, thus exploiting the embedding of the space of covariance operators in
the space of Hilbert--Schmidt operators, which is the infinite dimensional equivalent of embedding
covariance matrices in the space of symmetric matrices. However, \citet{pigoli2014distances} show
that better results can be achieved by using metrics that take into account the non Euclidean
geometry of the space of covariance operators. The drawback is that explicit analytic distributions
are not available for the test statistics based on these metrics and therefore the authors proposed
to use a permutation approach to carry out the test.

The aim of this work is to extend this idea to the case of multiple samples of functional data.
%However, most of the attention has been focused on the inference for the mean function (\citealp{fan1998test}, \citealp{cardot2003testing}, \citealp{cuevas2004ANOVA} and \citealp{shen2004f}).
%The problem of testing the equality of covariance operators, instead, has been studied only recently. \citet{panaretos2010second} derived a functional testing procedure under the assumption of Gaussianity in the two-sample case and \citet{fremdt2013testing} extended it to the non-Gaussian case.  In these works, the critical points of the testing procedures are obtained using asymptotic approximations of the test statistics distributions under the null hypothesis.
% Due to the complicated statistical functionals involved, the efficacy of these tests heavily rely on the accuracy of the approximations.
The testing of equality of several covariance operators has been first considered by
\citet{boente2014test}, that, in order to improve asymptotic approximations,  proposed to apply a
bootstrap procedure to calibrate the critical values of the test statistic obtained from the
Hilbert--Schmidt norm of the differences between sample covariance operators.
\citet{paparoditis2016bootstrap} investigated then the properties of an empirical bootstrap
methodology, applicable to more than two populations, but its consistency has been proven only for
test statistics based on the Hilbert--Schmidt norms and on the Karhunen--Lo\'{e}ve expansions of
the covariance operators. More recently, \citet{kashlak2016inference} applied concentration
inequalities to the analysis of covariance operators. These allow to construct non-asymptotic
confidence sets that can be used to make multiple-sample tests for the equality of covariances.

Since in the two-sample case the choice of the distance to define the test statistic has been shown
to impact the inferential performance in many scenarios \citep{pigoli2014distances}, we propose
here a more general approach that can be applied to test statistics defined through any valid
distance between covariance operators. Moreover, an appropriate choice of the permutation strategy
provides also pairwise tests between groups with a guaranteed control of the family-wise error
rate.

%Also, a permutation test for the equality of covariance operators in the two-sample case has been proposed by \citet{pigoli2014distances}: it can be used with any test statistic and makes no assumptions on the data except %for the exchangeability under the null hypothesis.

%The aim of this work is to extend the two-sample permutation test for covariance operators proposed by \citet{pigoli2014distances} to the multiple sample case.

Let us consider $q$ samples of random curves. We assume that curves in sample $i$:
\begin{equation}
 x_{i1}, \dots, x_{in_i} \in L^2(\Omega), \quad i = 1,\dots, q
\end{equation}
are realizations of a random process with mean $\mu_i$ and covariance operator $\Sigma_i$. We would
like to test the hypothesis
\begin{equation}
 H_0 : \{\Sigma_1 = \Sigma_2 =\dots = \Sigma_q\} \quad \text{ against } \quad H_1 :  \exists i \not = j \text{ s.t. } \Sigma_i \not = \Sigma_j.
\end{equation}
Moreover, if the null hypothesis $H_0$ is rejected, we would like to identify which pairs of groups
show a difference between covariance operators. To do this, we will rely on the non-parametric
combination methodology introduced by \citet{pesarin2010permutation} for multivariate permutation,
which enables to combine many different partial tests in an overall global test. In our case, the
idea is to combine all the pairwise comparisons between the $q$ samples in order to obtain the
$p$-value of the global test. Using this method, the post-hoc comparisons are straightforward: the
global $p$-value and the partial $p$-values  of the pairwise group comparisons are computed
simultaneously. However, some care is required when jointly analyzing the latter, because a
multiple testing problem arises. Thus, we suggest to use a step-down approach to control the
family-wise error rate. The empirical power of the proposed test is evaluated through simulation
studies and compared with those of previously proposed testing procedures. Finally, we analyze the
covariance operators of wheel-running mice activity curves. These data have been collected during
an evolutionary biology experiment to investigate the evolutionary behaviour of these activity
trait \citep[see][]{swallow1998artificial, koteja1999behaviour, kane2008basal}.
\section{Testing equality of covariance operators}
In this section, we describe the proposed strategy to test the equality of covariance operators
across multiple groups, which allows for the use of the most appropriate metric for covariance
operators in the problem at hand and, at the same time, for the investigation of pairwise
difference between groups. First, we discuss a few possible choices of distance between covariance
operators.

\subsection{Metrics for covariance operators}
\label{sec:metrics}
Let $x$ be a random function which takes values in $L^2(\Omega)$, $\Omega\subseteq \mathbb{R}$,
such that $E(||x||^2_{L^2(\Omega)})<+\infty$. The covariance operator $\Sigma_{x}$ is defined, for
$g \in L^2(\Omega)$, as $ \Sigma_{x} g(t)=\int_{\Omega}c_{x}(s,t)g(s)\mathrm{d}s, $ where
$c_{x}(s,t)=\mathrm{cov}(x(s),x(t))=E\left [ \left ( x(s)-E\left [ x(s) \right ]\right ) \left
( x(t)-E\left [ x(t) \right ]\right ) \right ].$ Then, $\Sigma_{x}$ is a trace class, self-adjoint,
compact operator on $L^2(\Omega)$ with non negative eigenvalues \citep[see, e.g., ][Section
1.5]{bosq2012linear}. Indeed, any compact operator $T$  has a canonical decomposition that implies
the existence of  two orthonormal bases $\{u_k\},\{v_k\}$ for $L^2(\Omega)$ such that $ Tf=\sum_k
\sigma_k \langle f,v_k\rangle u_k,$ or, equivalently, $ Tv_k=\sigma_k u_k, $ where $\langle
v,v\rangle$ indicates the inner product in $L^2(\Omega)$ and the non negative real numbers $\{\sigma_k\}_{k\in \mathbb{N}} $,  are called the singular values of $T$. If the operator is self-adjoint, there
exists an orthonormal basis $\{v_k\}$ such that $ Tf=\sum_k \lambda_k \langle f,v_k\rangle v_k,$
or, equivalently, $ Tv_k=\lambda_k v_k $ and the sequence $\{\lambda_k\}\in \mathbb{R}$ is called
the sequence of eigenvalues for $T$. A compact operator $T$ is said to be trace class if the trace
$ \mathrm{tr}(T)=\sum_k \langle Te_k,e_k \rangle <+\infty $ for every orthonormal basis $\{e_k\}$.
A compact operator $T$ is said instead to be Hilbert--Schmidt if its Hilbert--Schmidt norm is
bounded, i.e.,  $ ||T||^2_{\mathrm{HS}}=\mathrm{tr}(T'T)<+\infty$, where $T'$ denotes the adjoint operator of $T$. The Hilbert-Schmidt norm is a generalization of
the Frobenius norm for finite-dimensional matrices.

It is then possible to embed the space of covariance operators in the space of Hilbert--Schmidt
operators and use the Hilbert--Schmidt distance $||\Sigma_1- \Sigma_2||_{HS}$ to measure the
distance between two covariance operators $\Sigma_1$ and $\Sigma_2$. However, this is an extrinsic
metric based on the above embedding and thus ignores the geometry of the space of covariance
operators, such as the trace class property and the non negativity constraints on the eigenvalues.
\citet{pigoli2014distances} show that when the covariance operator is the object of interest for
the statistical analysis, taking into account the property of the space leads to tests with higher
empirical power. This motivated the introduction of new metrics such as the square root distance
and the Procrustes distance.

Let $\Sigma$ be a self-adjoint trace class operator, there exists a Hilbert-Schmidt self adjoint
operator
\begin{equation}
\label{eq:square} (\Sigma)^{1/2} f=\sum_k \lambda_k^{1/2} \langle f,v_k\rangle v_k,
\end{equation}
where $\lambda_k$ are eigenvalues and $v_k$ eigenfunctions of $\Sigma$. The square root distance
between two covariance operators $\Sigma_1$ and $\Sigma_2$ is therefore defined as
\begin{equation}\label{eq:sqdist}
d_R(\Sigma_1,\Sigma_2)=||\Sigma_1^{1/2}-\Sigma_2^{1/2}||_{\mathrm{HS}}.
\end{equation}

The square root distance is based on the mapping of the two operators $\Sigma_1$ and $\Sigma_2$ from the
space of covariance operators to the space of Hilbert--Schmidt operators, through the square root
map. This is a particular choice among a family of such maps that transform the covariance operator
$\Sigma$ to a Hilbert--Schmidt operator $L$ so that $\Sigma=L L'$. It is easy to see that $L$ is defined up to a unitary operator $R$, since $ (L
R)(LR)'=L R R' L'=L L'=\Sigma$. Therefore, it is natural to follow a Procrustes approach to
minimize the distance with respect to this arbitrary unitary operator. \citet{pigoli2014distances}
define the square of the Procrustes reflection size-and-shape distance between two covariance
operators  $\Sigma_1$ and $\Sigma_2$ as
\begin{equation}
\label{eq:proc} d_P(\Sigma_1,\Sigma_2)^2=\inf_{R\in O(L^2(\Omega))}||L_1-L_2
R||^2_{\mathrm{HS}}=\inf_{R\in O(L^2(\Omega))} \mathrm{trace}((L_1-L_2 R)'(L_1-L_2 R)),
\end{equation}
where $L_i$ are such that $\Sigma_i=L_i L_i'$, for $i=1,2$, and $O(L^2(\Omega))$ is the space of
unitary operators on $L^2(\Omega)$.

\subsection{Non-parametric combination}
\label{sec:multiple-sample}

In this section we describe how it is possible to test the global hypothesis that all the
covariance operators are equal across the groups by combining pairwise group comparisons which are
based on the two-sample permutation test described in  \citet{pigoli2014distances}. This approach
will allow us to use any metric in the definition of the test statistics without making any
assumption on the data generating process.

Let us assume we have $q$ independent groups of functional data
\begin{equation}
 x_{i1}, \dots, x_{in_i} \in L^2(\Omega), \quad i = 1, \dots, q.
\end{equation}
and they are independent and identically distributed samples from a random process with
distribution $P_i$, mean $\mu_i$ and covariance operator $\Sigma_i$. In the following, we denote with $\mathbf{x}_i$ the vector of observations $ (x_{i1}, \dots, x_{in_i})$ from group $i$.  We would like to test if the
covariance operators are all equal. The global null hypothesis can be viewed as an intersection of
partial null hypotheses and the global alternative hypothesis as the union of the corresponding
alternative hypotheses, i.e.
\begin{equation}
    H_0 : \bigcap\limits_{i \not = j} H_0^{ij} \text{ against } H_1 : \bigcup\limits_{i \not = j} H_1^{ij}, \text{ where } H_0^{ij} : \{\Sigma_i = \Sigma_j\}\text{ and } H_1^{ij} : \{\Sigma_i \not = \Sigma_j\}.
\end{equation}

The idea is to combine the $k=q(q-1)/2$ two-sample tests for each pair of groups in a global test,
using the non-parametric combination algorithm of \citet{pesarin2010permutation}. 

Let $T_{ij} = d(S_i, S_j)$ be the test statistic of our choice, associated to the partial test $H_{0}^{ij}$ of groups $i$ and $j$ respectively, where $S_i$, $S_j$ are sample covariance operators of the corresponding groups and $d(\cdot, \cdot)$ is some distance between covariance operators. In particular, in this work we consider the square root, Procrustes and Hilbert--Schmidt distances defined in Section \ref{sec:metrics}.
Let us define by ${\bf T} = (T_{1,2}, T_{1,3}, \dots, T_{q-1,q} )$, the vector of all partial test statistcs $T_{ij}$, with $1 \leq i <j \leq q$. 

The partial tests $H_0^{ij}: d(\Sigma_i,\Sigma_j) = 0$ against $H_1^{ij}: d(\Sigma_i,\Sigma_j) \not = 0$ marginally satisfy the assumptions required for the test (i.e. they are marginally unbiased, consistent and significant for large values) for any of the distances presented in \citet{pigoli2014distances}. Therefore, the considered algorithms can be applied to any functional dataset using the vector of test statistics ${\bf T}$. 

The partial test statistics in ${\bf T}$ are combined by a function $\Psi$ that must satisfy the properties indicated by \cite{pesarin2010permutation}:
\begin{enumerate}
 \item $\Psi$ is non-decreasing in each argument,%: $\Psi(\dotsc,\lambda_i,\dotsc) \geq \Psi(\dotsc,\lambda_i',\dotsc)$ if $\lambda_i\leq \lambda_i', i \in \{1, \dotsc, k\}$;
 \item If one or more arguments are zero, $\Psi$ attains its supremum value $\bar \Psi$, possibly not finite. %: $\Psi(\dotsc,\lambda_i,\dotsc) \to \bar \Psi$ if $\lambda_i\to 0, i=1,\dotsc,k$;
 \item For all $\alpha>0$, the critical value $T_{\Psi}^{\alpha}$ of $\Psi$ is assumed to be finite and strictly smaller than $\bar \Psi$.
\end{enumerate}

Also, the curves must be centred around the sample mean of each group, because exchangeability of the observations is required in order to apply permutations. 

We indicate by $x^{(0)}_{ij}$ the observations centred around the sample mean of the group $m_i$, by $\boldsymbol{x}_i^{(0)}$ the vector of centred observations of group $i$ and by $S^{(0)}_i$ the associated sample covariance operator. Similarly, we indicate by ${\bf u}^{(b)}$ the \emph{b}-th permutation of the data labels and so the superscript $(b)$ indicates the centred dataset, permuted according to ${\bf u}^{(b)}$.

We obtain the following algorithm:

\begin{myalgo}[Multiple-sample permutation test for the equality of covariance operators]\hfill \\
\label{algo:multiple-sample} Let $x_{ij}, i = 1, \dots, q, j = 1, \dots, n_i$ be the considered
dataset.
\begin{enumerate}
    \item Let $x_{ij}^{(0)} = x_{ij} - m_i$, where $m_i$ is the sample mean of ${\bf x}_i$, for all $i = 1, \dots, q$, $j = 1, \dots, n_i$.
    \item Let ${\bf T}^{(0)}$ be the $k$-dimensional vector containing the pairwise distances between the sample covariance operators of the centred groups ${\bf x}^{(0)}_i$ and ${\bf x}^{(0)}_j$,
        $d(S^{(0)}_i, S^{(0)}_j)$, for all $1 \leq i < j \leq q$.
    \item For $b = 1, \dots, B$, consider a random permutation ${\bf u}^{(b)}$ of the data labels and compute the $k$-dimensional vector ${\bf T}^{(b)}$ containing the distances between the sample covariance operators of the groups of the permuted data set, $d(S_i^{(b)}, S_j^{(b)})$, for all $1 \leq i < j \leq q$. $\{{\bf T}^{(b)}\}_{b=1}^B$ is a random sampling from the permutational distribution of the random vector ${\bf T}$.
    \item Let 
\begin{equation}
\hat{\lambda}_{ij} (d) = \frac{\sum_b \mathbbm{1} \big[d(S_i^{(b)}, S_j^{(b)})\geq d\big]}{B} 
\end{equation}    
    be consistent estimates of $\lambda_{ij}(d) = \mathbb{P} \big(d(S_i^{(b)}, S_j^{(b)}) \geq d \big)$, $d \in \mathbb{R}, d \geq 0$.
    \item Compute the estimated partial \emph{p}-values of the test as $\hat{\lambda}_{ij} = \hat{\lambda}_{ij}(d(S_i^{(0)}, S_j^{(0)}))$.
    \item Combine the $\hat{\lambda}_{ij}$  through the combining function $\Psi$ to obtain the observed global test statistic
    %\begin{equation}
        $T_{\Psi}^{(0)} = \Psi (\hat{\lambda}_{1,2}, \hat{\lambda}_{1,3},\dots, \hat{\lambda}_{q,q-1})$.
    %\end{equation}
    \item For $ b=1, \dots, B$, compute the \emph{b}-th combined test statistic as
    \begin{equation}
     T_{\Psi}^{(b)} = \Psi (\hat{\lambda}_{1,2}^{(b)},\hat{\lambda}_{1,3}^{(b)}, \dots, \hat{\lambda}_{q-1,q}^{(b)}), \text{ where }  \hat{\lambda}_{ij}^{(b)} = \hat{\lambda}_{ij}\big(d(S_i^{(b)},S_j^{(b)})\big).
    \end{equation}
    \item Compute the estimate of the \emph{p}-value of the combined test
        \begin{equation}
         \hat{\lambda}_{\Psi} = \frac{\sum_b \mathbbm{1} [T_{\Psi}^{(b)} \geq T_{\Psi}^{(0)}]}{B}.
        \end{equation}
    \item If $\hat{\lambda}_{\Psi} \leq \alpha$, reject $H_0$.
\end{enumerate}
\end{myalgo}
\begin{proposition}
If we make the additional assumptions that, when $n$ goes to infinity, then so also do the sample
sizes of all groups and that the number $B$ of Monte Carlo iterations goes to infinity while $k$ and
$\alpha$ remain fixed, then it is possible to prove that the test we obtain is strongly consistent
and unbiased for the overall null hypothesis $H_0$ against the alternative $H_1$.
\end{proposition}
This is a direct consequence of Theorems $2$, $4.3.1$ and $3$, $4.3.2$ of
\citet{pesarin2010permutation}.
\subsection{Synchronized permutation tests}
\label{sec:synchro}

When data belong to multiple groups, a few different strategies can be used to generate the permuted
samples. In particular, Step $3.$ of Algorithm \ref{algo:multiple-sample} requires to
generate a certain number of permutations of the original dataset. In \cite{solari2009permutation},
three different ways of permuting data are proposed. The simplest idea is to perform permutations
involving the whole data set, so-called pooled permutations.  This can be done because, under
$H_0$, the observations of all groups are exchangeable. However, this strategy does not allow to
test also the partial hypotheses, since each comparison involves not only the observations
belonging to the pair of considered groups, but also those of the other groups. Therefore, the
resulting global $p$-value is  correct, but the partial $p$-values would not be accurate when doing
post-hoc comparisons. The second proposal is to apply paired permutations, that is while comparing
the $i$-th and $j$-th groups, the inference is made on each paired vector $({\bf x}_i , {\bf x}_j)$
independently. The result would be opposite than the one obtained with pooled permutations: the
partial tests are exact, just like in the two-sample case, but the global test is not reliable
since this method does not take into account the dependencies between the marginal tests.

Therefore, we want paired permutations to be done not independently but jointly. At the same time,
we would like to keep the partial comparisons separate, so as to be able to do post-hoc comparison
with no additional computational effort. Then, if the design is balanced, i.e. $n_1 = \dots = n_q =
\bar{n}$, a further possibility explored by \cite{solari2009permutation} is to apply synchronized
permutations, exchanging the same number $\nu$ of units between each pair of blocks. Applying
synchronized permutations allows both maintaining the dependencies among partial tests and
involving the observations of each comparison at the same time. First of all, we build the
pseudo-data matrix
\begin{equation}\label{eq:pseudo-data-matrix}
\begin{bmatrix}
    \mathbf{x}_1 & \mathbf{x}_1 & \dots & \mathbf{x}_{q-1} \\
    \mathbf{x}_2 & \mathbf{x}_3 & \dots & \mathbf{x}_{q}
\end{bmatrix}
=
\begin{bmatrix}
   x_{1,1}     & x_{1,1}     & \dots & x_{q-1,1} \\
   x_{1,2}     & x_{1,2}     & \dots & x_{q-1,2} \\
   \vdots    & \vdots    &       &  \vdots   \\
   x_{1, \bar{n}}     & x_{1,\bar{n}}     & \dots & x_{q-1,\bar{n}}\\
             &           &       & \\
   x_{2,1}     & x_{3,1}     & \dots & x_{q,1}     \\
   x_{2,2}     & x_{3,2}     & \dots & x_{q,2}     \\
   \vdots    & \vdots    &       & \vdots    \\
   x_{2,\bar{n}}     & x_{3,\bar{n}}     & \dots & x_{q,\bar{n}}
\end{bmatrix}
\end{equation}
where each column is formed by the samples from two different groups and we have as many columns as
needed to account for all the groups pairings. Then, we can apply constrained synchronized
permutations, that is to exchange units in the same original position within each block. This can
be done by permuting the rows of the pseudo-data matrix. Since there are
\begin{equation}
    N_{\text{CSP}} = \binom{2\bar{n}}{\bar{n}}
\end{equation}
possible ways to exchange units in the first pair of blocks, $N_{\text{CSP}}$ is the cardinality of
the constrained synchronized permutations. Since the test statistic is symmetric, the number of
possible distinct values of the vector of test statistics ${\bf T}$ is $N_{\text{CSP}}/2$.

\subsection{Post-hoc analysis}
\label{sec:post-hoc} After performing the global test, if the null hypothesis $H_0$ is rejected in
favour of the alternative $H_1$, it is often of interest to find out which of the data samples led
to this conclusion. One of the advantages of the non-parametric combination methodology is that
partial $p$-values are computed at the same time of the global one. Therefore, the post-hoc
comparisons can be done with a small computational effort. We investigate here the methods that
allow to control the family-wise error rate, in order to simultaneously assess which of the partial
null hypotheses $H_0^{ij}$ are rejected.

First, we recall the resampling step-down method proposed by \cite{westfall1993resampling}.
The idea is that, rather than adjusting all $p$-values according to the minimum $p$-value distribution, one should only adjust
the minimum $p$-value using this distribution and then adjust the remaining $p$-values according to
smaller and smaller sets of $p$-values. The effect of using restricted sets of $p$-values is to
make the adjusted $p$-values smaller, thereby improving the power of the method.

Let the ordered partial $p$-values have indexes $r_1, \dotsc, r_k$ so that $\hat{\lambda}_{(1)}=\hat{\lambda}_{r_1}, \hat{\lambda}_{(2)}=\hat{\lambda}_{r_2}, \dotsc, \hat{\lambda}_{(k)}=\hat{\lambda}_{r_k}$. The step-down adjusted $p$-values are defined sequentially as follows:
\begin{equation}
    \tilde \lambda_{(1)} = \mathbb{P}\bigg(\min\limits_{j \in \{r_1, \dotsc, r_k\}} \hat \lambda_j \leq \hat\lambda_{(1)} | H_0 \bigg)
\end{equation}
\begin{equation}\label{eq:adjsd}
    \tilde \lambda_{(i)} = \max\bigg\{ \tilde\lambda_{(i-1)}, \mathbb{P}\bigg(\min\limits_{j \in \{r_i,\dotsc, r_k \}} \hat \lambda_j \leq \hat \lambda_{(i)}|H_0 \bigg) \bigg\}, \quad i = 2, \dotsc, k.
\end{equation}
The use of $\max$ operator insures that the order of the adjusted $p$-values is the same as that of
the original $p$-values. \cite{westfall1993resampling} proved that this procedure controls the
family-wise error rate in the strong sense.

\citet{pesarin2010permutation} showed that the resampling method proposed by
\cite{westfall1993resampling} is equivalent to iteratively use the non-parametric combination with
the Tippett combining function $\Psi_{\text{Tippett}}$ (\citealp{birnbaum1954combining}):
\begin{myalgo}[Step-down method for the Tippett combining function]\hfill
\label{algo:tippettstepdown}

    Let $\lambda_{(1)}, \dotsc, \lambda_{(k)}$ be the increasing ordered $p$-values corresponding to the set of partial hypotheses.
    \begin{enumerate}
        \item $\tilde \lambda_{(1)} = \Psi_{\text{Tippett}}(\lambda_{(1)},\dots,\lambda_{(k)}) = \min(\lambda_{(1)}, \dots, \lambda_{(k)}),$
        \begin{itemize}
            \item[--] If $\tilde \lambda_{(1)} \leq \alpha$, reject the corresponding hypothesis $H_{0}^{(1)}$ and continue;
            \item[--] Otherwise retain the hypotheses $H_{0}^{(1)}, \dotsc, H_{0}^{(k)}$ and stop.
        \end{itemize}
        \item For $i=2,\dotsc, k$, $\tilde \lambda_{(i)}= \Psi_{\text{Tippett}}(\lambda_{(i)},\dots,\lambda_{(k)})$
        \begin{itemize}
            \item[--] If $\tilde\lambda_{(i)}\leq \alpha$, reject also $H_{0}^{(i)}$ and continue;
            \item[--] Otherwise retain the hypotheses $H_{0}^{(i)}, \dotsc, H_{0}^{(k)}$ and stop.
        \end{itemize}
    \end{enumerate}
\end{myalgo}

Furthermore, \citet{lehmann2006testing} presented a similar step-down method, that uses the test statistics $T_{ij}$ instead of the $p$-values $\lambda_{ij}$. This method is equivalent to the one based on the Tippett combining function but allows to avoid the computations of the permutational distributions of the partial $p$-values. For example, let us suppose that the individual tests $H_{0}^{ij}$ are based on test statistics $T_{ij}$ with large values indicating evidence against the partial null hypotheses.  
Let $K = \{H_0^{ij}, 1 \leq i < j \leq q \}$ be the set of all the partial test hypotheses and $\bar K$ a subset of $K$, $\bar K \subseteq K$. First of all, we have to define the critical value of the combined test of all the hypotheses contained in $\bar K$ at level $\alpha \in [0, 1]$ so that the family-wise error rate is controlled in the strong sense. Many definitions are possible, as long as the properties indicated in \citet{lehmann2006testing}, Theorem 9.1.3 are verified. We choose to use the definition given in \citet{solari2009permutation}, where the critical value of $\bar K$ at level $\alpha$ is defined as the  $m$-th smallest value among the permutation distributions of $T_ {\bar{K}} = \max\limits_{H_0^{ij} \in \bar{K}} T_{ij}$
\begin{equation}
 c_{\bar K} (\alpha) =\bigg\{\max\limits_{ H_0^{ij} \in \bar{K}} T_{ij}^{(b)}, b = 1, \dotsc, B \bigg\}_{(m)},
\end{equation}
where $m = B - \lfloor B \alpha \rfloor$. For this reason we will refer to this as the step-down method for the $\max T$ combining function. The algortithm is defined as follows:

\begin{myalgo}[Step-down method for the $\max T$ combining function]\hfill
\label{algo:genericstep-down}

	Let $ T_{(1)} = T_{r_1} \geq \dotsc \geq T_{(k)} = T_{r_k} $ denote the observed ordered test statistics where ${r_1, \dotsc, r_k}$ are such that $T_{r_1} \geq T_{r_2} \geq \dotsc \geq T_{r_k}$ and let $H_0^{(1)}, H_0^{(2)}, \dotsc, H_0^{(k)}$ be the corresponding hypotheses.
	\begin{enumerate}
		\item Let $K_1 = K$,
		\begin{itemize}
			\item[--] If $ T_{r_1} \geq c_{K_1}(\alpha)$ reject $H_0^{(1)}$ and continue;
			\item[--] Otherwise retain the hypotheses $H_{0}^{(1)}, \dotsc, H_{0}^{(k)}$ and stop.
		\end{itemize}
		\item For $i=2,\dotsc, k$, let $K_i$ be the set of hypotheses not previously rejected, i.e. $K_i = \{H_0^{(i)}, \dotsc, H_0^{(k)}\}$,
		\begin{itemize}
			\item[--] If $T_{r_i} \geq c_{K_i}(\alpha)$ reject $H_0^{(i)}$ and continue;
			\item[--] Otherwise retain the hypotheses $H_0^{(i)}, \dotsc, H_0^{(k)}$ and stop.
		\end{itemize}
	\end{enumerate}
\end{myalgo}

Lastly, when using another combining function, it is possible to use the closed testing procedure
of \citet{marcus1976closed}. This method is based on the idea that one may reject any hypothesis
$H_{0}^{ij}$, while controlling the family-wise error rate, when the test of $H_{0}^{ij}$ itself is
significant and the test of every intersection of partial hypotheses that includes $H_{0}^{ij}$ is significant.
Hence, $ \tilde{\lambda}_{ij}$ is the maximum of all the $p$-values of the partial hypotheses containing $H_0^{ij}$. This method has two
major drawbacks: it requires a greater number of computations and it is very conservative. However,
it is a useful tool when the use of the Tippett or $\max T$ combining functions is not suitable.
\section{Simulation studies}
\subsection{Synthetic data sets}
\label{sec:casestudies} We generate synthetic data sets as follows. All the curves are generated on
an equispaced grid of $p = 31$ points on $ \Omega = [0,1]$ and the sample size of each group is $\bar{n} = 20$.
Unless otherwise stated, curves are simulated from a multivariate Gaussian process. We consider $q$
different groups (with $q$ varying across simulation studies) and for all $q$ groups the mean
function is equal to $\sin(x)$, $x \in [0,1]$. The covariance operator of each group varies
according to the test case. Let $\Sigma_1$ and $\Sigma_2$ be the sample covariance operators of the
male and female subjects in the Berkeley growth study dataset described in
\citet{ramsay2005functional}, rescaled to $[0,1]$. 

\begin{figure}
\centering
%\hspace{-4cm}
\begin{subfigure}{.3\textwidth}
\begin{tikzpicture}[scale=0.7]
	\begin{axis}[colormap/bluered,mesh/ordering=x varies,mesh/cols=31,point meta min=5, point meta max=80, zmin=5,zmax=80,view={-45}{65}]
		\addplot3[surf,x=x,y=y,z=vec10] table {tikz/C10.txt};% table {\datatablePX};
	\end{axis}
\end{tikzpicture}
\caption{Male ($\Sigma_1$).}
\end{subfigure}

%\hspace{-4cm}
\begin{subfigure}{.3\textwidth}
\begin{tikzpicture}[scale=0.7]
	\begin{axis}[colormap/bluered,mesh/ordering=x varies,mesh/cols=31,point meta min=5, point meta max=80, zmin=5,zmax=80,view={-45}{65}]
		\addplot3[surf,x=x,y=y,z=vec10] table {tikz/CF.txt};% table {\datatablePX};
	\end{axis}
\end{tikzpicture}
\caption{Female ($\Sigma_2$).}
\end{subfigure}
\caption{Covariance operators of the subjects in the Berkeley growth study dataset.}
\end{figure}
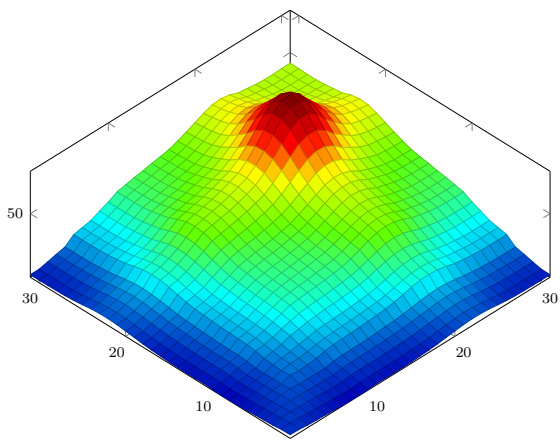
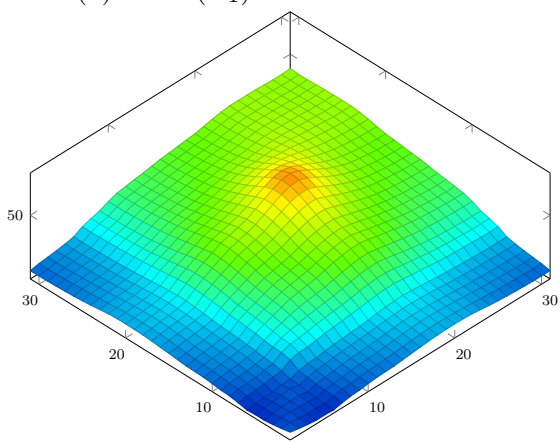

We consider two forms for the expression of the covariance operators of some of the groups under the alternative:
\begin{description}
\item[First test case] Some of the groups have covariance operator $ \Sigma(\gamma)=[(\Sigma_1)^{1/2}+\gamma\{(\Sigma_2)^{1/2}\tilde{R}-(\Sigma_1)^{1/2}\}][(\Sigma_1)^{1/2}+\gamma\{(\Sigma_2)^{1/2}\tilde{R}-(\Sigma_1)^{1/2}\}]'$
where $\tilde{R}$ is the operator minimizing the Procrustes distance between $\Sigma_1$ and
$\Sigma_2$ (\citealp{pigoli2014distances}) and $\gamma\in [0,5] $ is a parameter which controls how
far this covariance operator is from $\Sigma_1$.

\item[Second test case] Some of the groups have covariance operator  $ \Sigma(\gamma)=(1+\gamma)\Sigma_1$, $\gamma\in [0,5]$.
\end{description}

The two test cases represent two different ways in which the null hypothesis can be violated. The
second case pertains to a difference in the total variation between groups, while the first
test case presents also a difference in shape between covariance operators.

Each permutation test is performed with $B = 1000$ iterations of the Monte Carlo Algorithm
\ref{algo:multiple-sample} and is repeated for $1000$ replicates of the simulated dataset.

In the following, we use this simulated data to evaluate the empirical size and the empirical power
of the proposed test when using different distances between covariance operators.

All the functions needed to apply the permutation tests to these simulated data have been made
available in the \textsf{R} package ``fdcov'' \citep{cabassi2016fdcov}.

\subsection{Empirical size and power of the test}

\begin{figure}
    \centering
    \begin{subfigure}{.48\textwidth}\includegraphics[width=\textwidth]{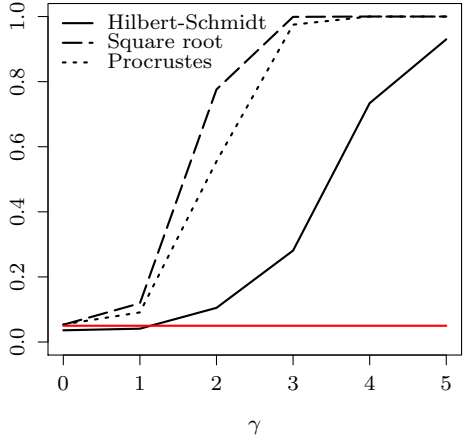}\caption{Global test.}\end{subfigure}
    \begin{subfigure}{.48\textwidth}\includegraphics[width=\textwidth]{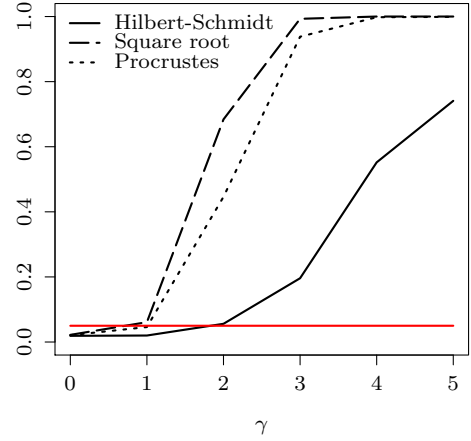}\caption{Samples 1 and 2.}\end{subfigure}
    \begin{subfigure}{.48\textwidth}\includegraphics[width=\textwidth]{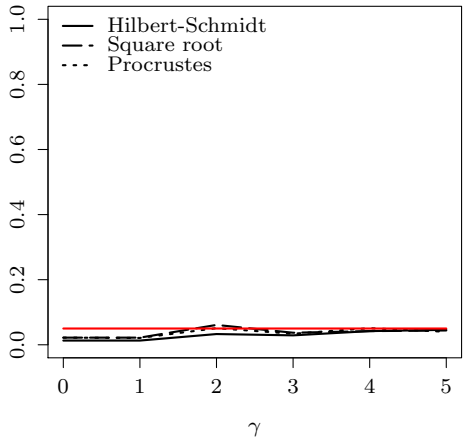}\caption{Samples 2 and 3.}\end{subfigure}
    \begin{subfigure}{.48\textwidth}\includegraphics[width=\textwidth]{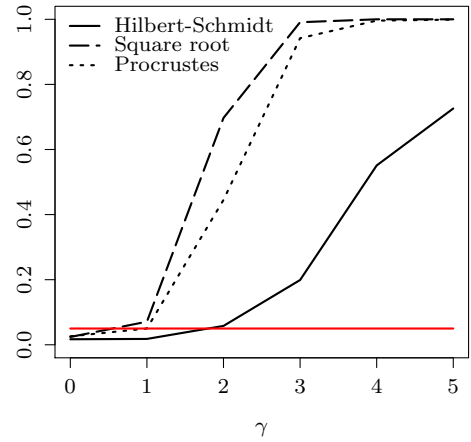}\caption{Samples 1 and 3.}\end{subfigure}
    \caption{Empirical power of synchronized permutation global and partial tests applied to the first test case using $\max T$ combining function. $p$-values have been adjusted using the $\max T$ step-down procedure.}
    \label{fig:max1}
\end{figure}

\begin{figure}[t!]
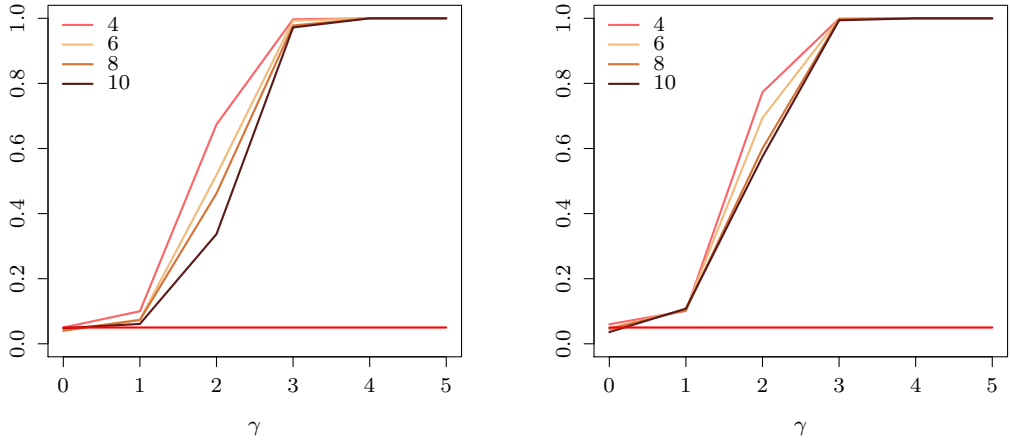

    \centering
    \begin{subfigure}{.48\textwidth}\includegraphics[width=\textwidth]{tikz/more_one.tikz}\caption{One group has fixed covariance $\Sigma_1$, the others have covariance  $\Sigma_{\gamma}$.}\end{subfigure}
    \begin{subfigure}{.48\textwidth}\includegraphics[width=\textwidth]{tikz/more_half_new.tikz}\caption{Half of the groups have fixed covariance $\Sigma_1$, the others have covariance $\Sigma_{\gamma}$.}\end{subfigure}
    \caption{Empirical power of synchronized permutation global tests applied to the first test case using the $\max T$ combining function, with 4, 6, 8 and 10 data samples.}
    \label{fig:more}
 \end{figure}

We consider first a simulation with $q=3$ groups, where the first group has covariance operator
$\Sigma_1$ and the others two covariance operators $\Sigma(\gamma)$. Figures \ref{fig:max1} and
\ref{fig:max2} show the empirical power of the global and partial tests done using the synchronized
permutations, the $\max T$ combining function and the Procrustes, square root and Hilbert--Schmidt
distance, for the first and second test cases respectively. It is evident that the test has greater
empirical power when using Procrustes and square root distances, with the latter being in this case
preferable due to the lower computational cost. Moreover, the global test appears to have the
correct level for all the distances while the partial tests are conservative for $\gamma=0$, as
expected, and the proportion of rejection for the partial test between the second and third group
(which have equal covariances) is less than $5\%$ for all values of $\gamma$.
We want now to explore how the performance of the test changes when the number of groups increases.
Figure \ref{fig:more}(a) shows the empirical power of the global test using the square root
distance when the number of groups goes from $4$ to $10$, always with the first group with
covariance operator $\Sigma_1$ and all the others with covariance operator $\Sigma(\gamma)$, with
$\gamma$ varying from $1$ to $5$. It is possible to see that the level of the test is respected for
all numbers of groups while the empirical power tends to decrease when the number of groups
increases. This is due to the fact that only $q$ partial tests out of $q(q-1)/2$ are bringing information about the
violation of the null and they form a smaller and smaller proportion of all the partial tests when
$q$ increases. If we instead have half of the groups with covariance operator $\Sigma_1$ and half
with covariance operator $\Sigma(\gamma)$, the loss of empirical power when $q$ increases is smaller, as shown by the empirical power curves reported in Figure \ref{fig:more}(b). This is because of the larger proportion of false partial hypothesis.
\begin{figure}
    \begin{subfigure}{.48\textwidth}\includegraphics[width=\textwidth]{tikz/GLadj_dist2.tikz}\caption{Global test.}\end{subfigure}
    \begin{subfigure}{.48\textwidth}\includegraphics[width=\textwidth]{tikz/ABadj_dist2.tikz}\caption{Samples 1 and 2.}\end{subfigure}
    \begin{subfigure}{.48\textwidth}\includegraphics[width=\textwidth]{tikz/BCadj_dist2.tikz}\caption{Samples 2 and 3.}\end{subfigure}
    \begin{subfigure}{.48\textwidth}\includegraphics[width=\textwidth]{tikz/ACadj_dist2.tikz}\caption{Samples 1 and 3.}\end{subfigure}
    \caption{Empirical power of synchronized permutation global and partial tests applied to the second test case using the $\max T$ combining function. $p$-values have been adjusted using the $\max T$ step-down procedure.}
    \label{fig:max2}
\end{figure}
%
%\begin{figure}
%    \centering
%    \begin{subfigure}{.48\textwidth}\includegraphics[width=\textwidth]{tikz/Sim/maxT/Sq1.tikz}\caption{First test case.}\end{subfigure}
%    \begin{subfigure}{.48\textwidth}\includegraphics[width=\textwidth]{tikz/Sim/maxT/Sq2.tikz}\caption{Second test case.}\end{subfigure}
%    \caption{Empirical power of synchronized permutation global tests applied to the first and second test case using $\max T$ combining function and the three different types of permutations: pooled, paired and synchronized.}
%    \label{fig:perm}
% \end{figure}
 %

\subsection{Comparison with the other existing tests}
We compare now the proposed method, using the square root distance and the $\max T$ combining
function, with some alternative approaches to test for the difference between covariance operators.
We consider first a generalization of the Levene's test (\citealp{anderson2006distance}) which is
sensitive  only to the difference in total variation between groups and it is implemented using the
permutational analysis of variance. \citet{paparoditis2016bootstrap} introduced an empirical
bootstrap approach based on Hilbert--Schmidt distance (or alternatively, on other test statistics
based on the Karhunen--Lo\'{e}ve expansions of the covariance operators). In the interest of a fair
comparison, we apply here the same procedure to the test statistics based on the square root
distance. It should be noted however that the theoretical properties of this modified procedure
still need to be studied. Finally, we consider the test based on the concentration inequalities
method of \citet{kashlak2016inference}.

\begin{figure}[!t]
\centering
    \begin{subfigure}{.45\textwidth}\includegraphics[width=\textwidth]{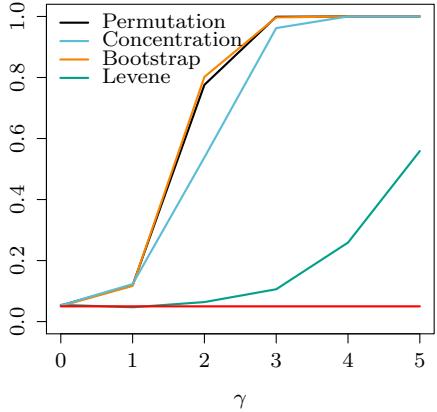}\caption{First test case.}\end{subfigure}
    \begin{subfigure}{.45\textwidth}\includegraphics[width=\textwidth]{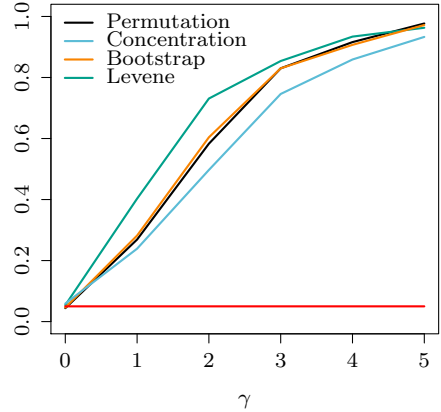}\caption{Second test case.}\end{subfigure}
    \caption{Empirical power of synchronized permutation, Levene's, empirical bootstrap and concentration inequalities-based global tests applied to the first (left) and second (right) test cases.
    Data are sampled from a Gaussian process.
    The results shown were obtained using the combining function $\max T$ and the $p$-values have been adjusted with the step-down procedure.}
    \label{fig:other_gauss}
\end{figure}

Figures \ref{fig:other_gauss} show the empirical power of the generalisation of Levene's test
(\citealp{anderson2006distance}), the empirical bootstrap by \citet{paparoditis2016bootstrap} and
the concentration inequalities method of \citet{kashlak2016inference} for the two test cases,
compared to the results obtained using the proposed permutation test. Here data are simulated from
$q=3$ groups with the first group having covariance operator $\Sigma_1$ and the other two
covariance operators $\Sigma(\gamma)$. It appears that the permutation test and the empirical
bootstrap have approximately the same empirical power in both test cases. On the contrary,
Levene’s test performs very differently. As expected, it outperforms the others in the second
test case, where it captures very well the differences in scale, but it is dramatically less
powerful in the first test case, where the difference between the covariance operator is mostly in
shape. The non-asymptotic test of \citet{kashlak2016inference} is slightly less powerful than the
permutation test and the empirical bootstrap but it has the advantage of being much less
computationally expensive than the resampling-based methods.

\begin{figure}[!h]
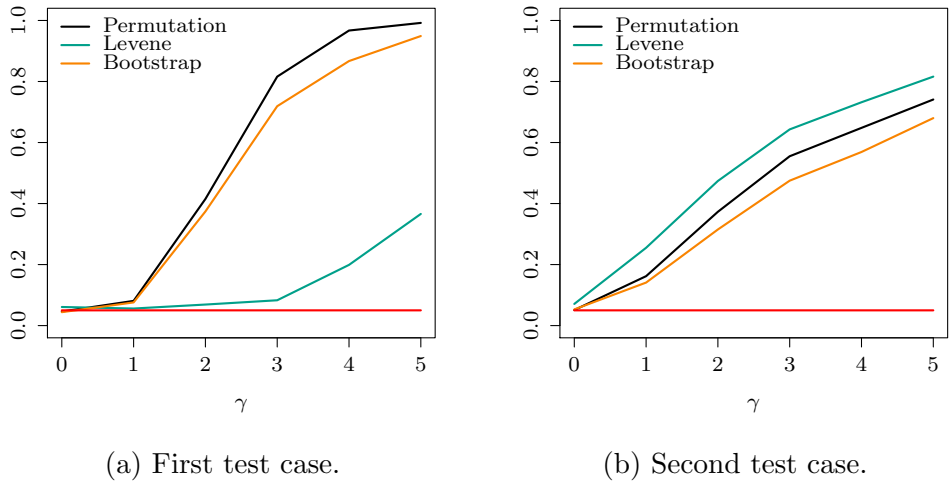

\centering
    \begin{subfigure}{.45\textwidth}\includegraphics[width=\textwidth]{tikz/tStud1.tikz}\caption{First test case.}\end{subfigure}
    \begin{subfigure}{.45\textwidth}\includegraphics[width=\textwidth]{tikz/tStud2.tikz}\caption{Second test case.}\end{subfigure}
    \caption{Empirical power of synchronized permutation, Levene's and empirical bootstrap global tests applied to the first (left) and second (right) test cases.
    Data are generated using a multivariate t-Student.
    The results shown were obtained using the combining function $\max T$ and the $p$-values have been adjusted with the step-down procedure.}
    \label{fig:other_nongauss}
\end{figure}

We want also to explore what happens when data are generated from a non Gaussian distribution. We
simulated data from a multivariate $t$ distribution with $4$ degrees of freedom and correlation
matrix implied by the covariance operator $\Sigma_1$ for the first group and $\Sigma(\gamma)$ for
the other two groups. Here it is not possible to apply the non-asymptotic test of
\citet{kashlak2016inference}, because calibration parameters are not yet available when data are
not Gaussian. Figure \ref{fig:other_nongauss} shows the empirical power for the permutation test,
the empirical bootstrap test and the Generalized Levene's test. Here the permutation test appears
to perform slightly better then the bootstrap, while Levene's test is again performing very well in
the second tests case but not in the first. Overall, the empirical power of all tests is lower than
in the Gaussian case, but they respect the nominal level.

\section{Application to evolutionary biology}

In this section we apply the proposed permutation test to a data set of interest in evolutionary
biology. The question of interest here is whether there is a difference in the covariance operator
of a function-valued trait \citep[][]{kingsolver2001variation, stinchcombe2012genetics} among
experimental lines of mice with known differences in evolutionary histories.

\subsection{Data set}

Data were collected from aging house mice (\emph{Mus domesticus}) that were members of the $16$th
generation of a selective breeding experiment for increased voluntary wheel-running behavior
\citep{swallow1998artificial}. This experiment produced four replicate lines selected for the total
number of wheel revolutions run on days $5$ and $6$ of a six day exposure to running wheels that
occurred when the mice were six to eight weeks of age, and four replicate control lines that were
randomly bred each generation \citep[see][for additional details]{swallow1998artificial}. At
generation $16$, a total of $360$ mice were used to establish an aging colony
\citep{morgan2003ontogenies,bronikowski2006evolution}. Half of the mice in the colony were from the
four high-selected lines and half were from the four control lines that were randomly bred with
respect to running behavior, and half of each selection group was housed with running wheels
(active mice) and half was housed in cages without wheels (sedentary mice). 
%Approximately half of
%each group (selective active, selective sedentary, control active, control sedentary) were
%sacrificed after 80 weeks; the rest of the individuals were allowed to live out their lifespan. 
One male from one of the control lines died of unknown causes during the early stages of the
experiment.  Each week every mouse was measured for body mass and food consumed, and each active
mouse had the total number of wheel revolutions run that week recorded \citep[see][for more
details]{morgan2003ontogenies,bronikowski2006evolution}.

Herein we examine only data from the active mice from both selection groups from the first $80$
weeks of the experiment, as reported by \citet{morgan2003ontogenies}.  The variables in the dataset
are: a unique id number for each mouse and id of fullsib family from which the mouse was drawn; the
age and sex of the mouse; the line number (lines 1, 2, 4, 5 are control lines, the others are
selected lines); the week of wheel measure and the number of revolutions run during the week. Total
activity, measured as number of revolutions run in a given day, can be decomposed into the product
of mean velocity and duration of activity. Thus, increased total activity levels could be
accomplished by an increase in mean velocity, an increase in the amount of time spent running, or a
combination of both.

\begin{figure}[h!]
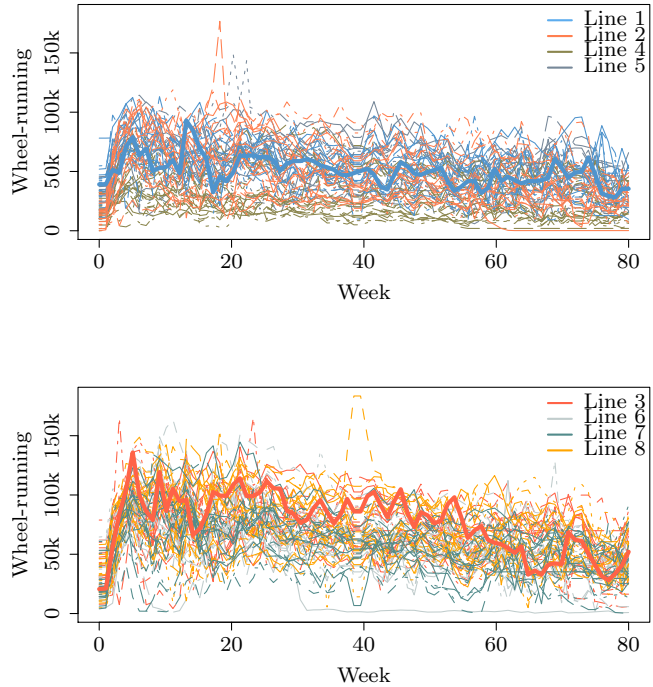

\centering
    \begin{subfigure}{.6\textwidth}
        \includegraphics[width=\textwidth]{tikz/mice_control80.tikz}
%       \caption{Control lines}
    \end{subfigure}
    \begin{subfigure}{.6\textwidth}
        \includegraphics[width=\textwidth]{tikz/mice_selected80.tikz}
%       \caption{Selected lines}
    \end{subfigure}
    \caption{Voluntary wheel running activity dataset, raw data, control and selected lines.}
    \label{fig:rawdata}
\end{figure}

The raw data are presented in Figure \ref{fig:rawdata}. Each line connects the number of
revolutions run by each mouse during the first $80$ weeks of the experiment
\citep{morgan2003ontogenies}. Mice identified by ID numbers 90183 and 90224 are taken as examples
of the selected and control lines respectively. The corresponding wheel-running functions have been
highlighted in each figure. The first one is a male belonging to family number $29$ from line $1$
(control), while the second one is a female belonging to family number $11$ from line $3$
(selected).

At several times during the experiment, data collection was skipped for one or two weeks. In these
cases, the data collected after the skipped week(s) was divided by number of weeks, giving multiple
weeks in a row with the same value. This is easily seen in Figure \ref{fig:rawdata} at weeks $38$,
$39$, $40$, when the values are constant for each mouse, because the wheel revolutions recorded for
week $40$ were divided by 3 and assigned to weeks 38 and 39 as well as 40. The weeks in which this
occurred are: $34$; $35$; $38$; $39$; $40$; $50$; $51$; $72$; $73$.

\begin{figure}[h!]
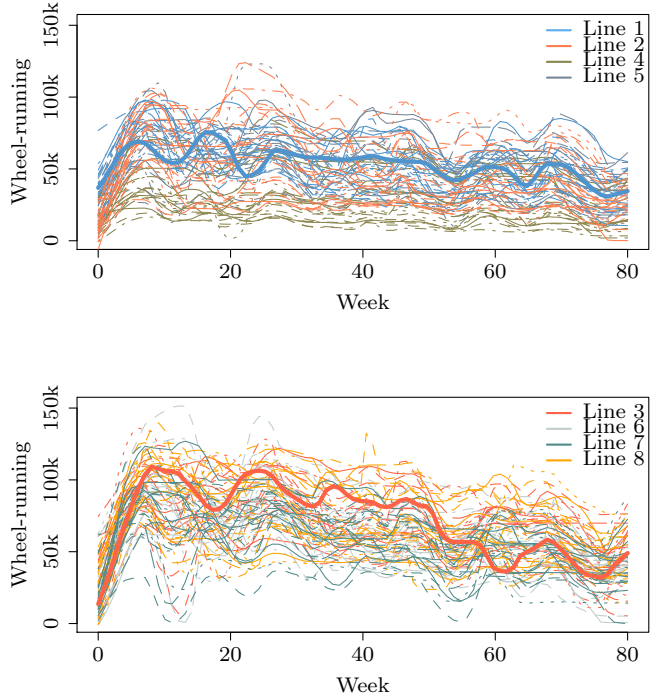

\captionsetup[subfigure]{justification=centering} \centering
    \begin{subfigure}{.6\linewidth}
    \includegraphics[width=\textwidth]{tikz/mice_control80align.tikz}
%   \caption{Control lines}
    \end{subfigure}\\
    \begin{subfigure}{.6\linewidth}
    \includegraphics[width=\textwidth]{tikz/mice_selected80align.tikz}
%   \caption{Selected lines}
    \end{subfigure}
    \caption{Voluntary wheel running activity dataset after smoothing and alignment, control and selected lines.}
    \label{fig:alignment}
\end{figure}

We regularized data using cubic smoothing splines. In particular, we used the the routine
\verb+spline.smooth()+ of the \textsf{R} package  ``stats'' \citep{rcore2016r}.
%The breakpoints correspond to the times of data collection. We have assigned null weight to all the weeks when the number of revolutions has not been recorded correctly and equal to one for all the others.
%Applying it to the evolutionary data set, we obtain the results presented in Figure \ref{fig:smoothing}.
%\begin{figure}
%\captionsetup[subfigure]{justification=centering}
%\centering
%    \begin{subfigure}{.6\linewidth}
%    \includegraphics[width=\textwidth]{tikz/Mice/c80s.tikz}
%%   \caption{Control lines}
%    \end{subfigure}\\
%    \begin{subfigure}{.6\linewidth}
%    \includegraphics[width=\textwidth]{tikz/Mice/s80s.tikz}
%%   \caption{Selected lines}
%    \end{subfigure}
%    \caption{Evolutionary biology data set after smoothing, control and selected lines.}
%    \label{fig:smoothing}
%\end{figure}
%%
Since individual mice can have their own biological clock, curves are aligned to remove phase
variability (\citealp{ramsay2005functional}), via the elastic analysis described in
\citet{tucker2013generative} and implemented in the \textsf{R} package ``fdasrvf''
\citep{tucker2016fdasrvf}. Figure \ref{fig:alignment} shows the smooth and aligned wheel-running
activity curves.
\subsection{Missing observation}
In the voluntary wheel running activity data set, all groups  (experimental lines) are composed of
$20$ mice. However, one of the mice died of unknown causes during the early stages of the
experiment and therefore one group has only $19$ observations. For this reason, in order to apply
the synchronized permutations, we have to prove that the presence of a missing observation does not
affect the inference. Following the guidelines given by \citet{pesarin2010permutation}, we give a
new formulation of the test that takes into account the presence of missing data. Thanks to this,
we are able to prove that it is possible to apply the proposed test to an unbalanced data set with
one missing observation, under certain assumptions on the process that generates the missing
observations.

Consider again a functional data set of the form
\begin{equation}
{\bf X} =\{x_{ij},\ i=1,\dots,q,\ j=1,\dots,n_i\}, \label{eq:dataset}
\end{equation}
that consists of $q \geq 2$ samples of size $n_i \geq 2$. The groups are
related to $q$ levels of a treatment and the data $x_{ij}$ are supposed to be independent and
identically distributed with distributions $P_i \in \mathcal{P}$, $i = 1, \dots,q$. In order to
take into account that, for whatever reason, some of the data are missing,
\citet{pesarin2010permutation} suggested to consider the inclusion indicator associated to the
considered data set, that is
\begin{equation}
{\bf O} =\{o_{ij},\ i = 1,\dots,q,\ j = 1,\dots, n_i\},
\end{equation}
where $o_{ij}=1$ if $x_{ij}$ has been observed and collected, $o_{ij}=0$ otherwise. We denote with ${\bf o}_i$ the vector of observation indicators $(o_{i1}, \dots, o_{in_i})$ from group $i$. This indicator
represents the observed configuration in the data set. Hence, the data set can be seen as the pair
of matrices $({\bf X},{\bf O})$.
% We define by $\kappa_i=\sum_j O_{ij}$ the actual sample size of the observed data of each sample.
Therefore we would like to perform the following test:
\begin{equation}
 H_0 : \{ ({\bf x}_1,{\bf o}_1) \overset{d}{=} \dots \overset{d}{=} ({\bf x}_q,{\bf o}_q)\} \quad \text{against} \quad H_1 : \{H_0 \text{ is not true}\}.
\end{equation}
Thus, if we assume that data are jointly exchangeable under the null hypothesis with respect to the
groups, we can, again, utilize a permutation test. Let us represent by $P_i$ the joint multivariate
distribution of $({\bf x}_i,{\bf o}_i),i = 1,\dots,q$ under the null hypothesis. Then it holds:
\begin{equation}
P_i = P_{{\bf o}_i} \cdot P_{{{\bf x}_i}|{{\bf o}_i}}.
\end{equation}
The idea of \citet{pesarin2010permutation} is to break down the null hypothesis in the following
way:
\begin{equation}
 H_0 : \{ [{\bf o}_1 \overset{d}{=} \dots \overset{d}{=} {\bf o}_q] \cap [{\bf x}_1 \overset{d}{=} \dots \overset{d}{=} {\bf x}_q | {\bf O}] \} = \{H_0^{\bf O} \cap H_0^{{\bf X}|{\bf O}}\}.
\end{equation}
Furthermore, we assume that the missing data are missing completely at random. In this case, we can
condition with respect to the observed inclusion indicator and ignore $H_0^{\bf O}$, because ${\bf
O}$ does not provide any information about treatment effects (\citealp{rubin1976inference}). In
other words, the partial hypotheses on ${\bf O}$ are true by assumption and the null hypothesis can be
simplified:
\begin{equation}
 H_0 = H_0^{{\bf X}|{\bf O}} = \{{\bf x}_1 \overset{d}{=} \dots \overset{d}{=} {\bf x}_q | {\bf O} \}.
\end{equation}
We indicate by ${\bf O}^*$ any permutation of ${\bf O}$, the permutational vector of inclusion
indicators, and by $\boldsymbol{\kappa}^* = [\kappa^*_1, \dots, \kappa^*_q]$ the corresponding vector of counts of valid observations in each group,
where
\begin{equation}
\kappa^*_i = \sum_{j=1}^{n_i} o^*_{ij}, i = 1, \dots, q.
\end{equation}
Then we can group the set of all permutations of the dataset, according to the vectors of actual
sample sizes of valid data $\boldsymbol{\kappa}^*$. Now, let  ${\bf T}$ be the vector of partial
test statistics based on functions of sample valid data; we denote its permutation distribution as
$F[{\bf T}|({\bf X},{\bf O})]$, ${\bf T} \in \mathbb{R}^k$. \citet{pesarin2010permutation}  pointed
out that, if the permutation sub-distributions of the partial test statistics are invariant with
respect to the sub-groups induced by ${\bf O}^*$,  then we can simply evaluate $F[{\bf T}|({\bf
X},{\bf O})]$ ignoring the missing values. This implies that
\begin{equation}
 F[{\bf T} | ({\bf X},{\bf O})] = F[{\bf T} | ({\bf X}, {\bf O}^*)]
\end{equation}
holds for every ${\bf T} \in \mathbb{R}^k$, for every permutation ${\bf O}^*$ of ${\bf O}$ and for
all data sets ${\bf X}$. In the case of the tests for covariance operators, this is true because
the test statistic $T_{\Psi}$ of the global test is a combination of the partial test statistics of the
pairwise comparisons between the groups. These, in turn, depend only on the distances between
covariance operators and their permutations. We can suppose that, under the null hypothesis, the
permutation distribution of the partial test statics $T_{ij}$ depends essentially on the number $\kappa^*_i, \kappa^*_j$ of summands. Thus, just like in the case of the multivariate analysis of variance
studied in \citet{pesarin2010permutation}, the previous distributional equality is equivalent to
\begin{equation}
 F[{\bf T}| ({\bf X},\boldsymbol{\kappa})] = F[{\bf T}| ({\bf X},\boldsymbol{\kappa}^*)],\label{eq:distrib-equality}.
\end{equation}
Hence, we would like our partial test statistics to be invariant with respect to
$\boldsymbol{\kappa}^*$ and for all ${\bf X}$. Now, suppose that we are in the balanced case, i.e.
$n_1 = \dots = n_q = \bar{n}$ and one observation is missing in one of the groups, say group $a$,
where $1 \leq a \leq q$. In the wheel-running data set, for instance, $q=8$, $\bar{n}=20$ and one
observation is missing in group $1$. All the pairwise comparisons between groups $i$ and $j$ with $1 \leq i < j \leq q$ and $i,j \not = a$ are not affected by the
problem of missing data since $\kappa_i^*=\kappa_j^*=\bar{n}$. As regarding the others, at each
iteration of the algorithm, we could have $\kappa_a^* = \bar{n}$ and $\kappa_j^*=\bar{n}-1$ or
viceversa, depending on the permutation. However, since distances are symmetric, this two cases are
permutationally equivalent under the null hypothesis and Equation \eqref{eq:distrib-equality} is
always satisfied. For this reason, we can apply the synchronised permutations as usual. At each
iteration of Algorithm \ref{algo:multiple-sample} the sample covariance of each permuted group is
computed only with the available data. This is more complicated when the number of missing data
becomes greater than one, since the vector $\boldsymbol{\kappa}^*$ of actual sample sizes can assume other
values.
\subsection{Hypothesis testing}
We can finally apply the test to the smoothed and aligned wheel-running activity curves. The aim of
the analysis is to check if the covariance operators of the eight groups of mice are the same and,
if this is not the case, to identify which lines have different covariances. This is necessary for two reasons. First, the covariance operator is in itself of biological interest for exploring which
type of variability is environmental in nature and which is due to genetic components. Second,
inference on the mean functions often requires the assumption of equality of covariance operator
and it is important to be able to check this assumption.

We want then to test the hypothesis
\begin{equation}
 H_0: \{\Sigma_1 = \dots = \Sigma_8\} \quad \text{against} \quad H_1 : \{\text{at least one of the equalities is not true}\}.
\end{equation}
To this end, we use the Monte Carlo Algorithm \ref{algo:multiple-sample} to obtain an estimate of
the permutation test proposed in Section \ref{sec:multiple-sample}. We have shown in the previous
section that synchronized permutations can be used, even if one of the observations is missing. We
use here the square root distance between covariance operators as partial test statistic and we
choose the Tippett combining function. We set the number of iterations $B$ to $1000$.

\begin{figure}[h!]
\centering
    \includegraphics[width=.6\textwidth]{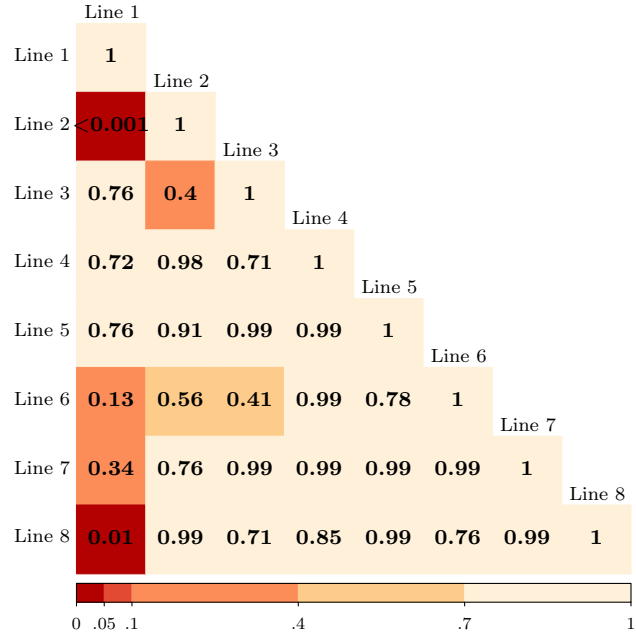}
    \caption{Partial $p$-values of the synchronized permutation test on the covariance operators of the aligned data. For each $1 \leq j \leq i \leq q$, $i \not = j$, the value reported in row $i$, column $j$ corresponds to the adjusted $p$-value of the test $H_0: \{\Sigma_i = \Sigma_j \}$ against $H_1:\{\Sigma_i \not = \Sigma_j\}$. The global $p$-value of the test is the minimum of the partial $p$-values and therefore is less than 0.001.}
    \label{fig:covpvalues}
\end{figure}

The $p$-values of the partial tests between each pair of lines, adjusted with the step-down method,
are reported in Figure \ref{fig:covpvalues}. The $p$-value of the global test ($<0.001$) indicates
that there is strong evidence to reject the null hypothesis. This is due mainly to the first group
of mice for which many partial null hypotheses are rejected (i.e., the differences between the
covariance operator of line $1$ and the covariance operators of these others lines are significant)
and, when using the $\max T$ combining function, we reject $H_0$ even if only one of the partial tests
is rejected.

The results of this test are somewhat surprising.  First, no systematic differences were detected
between the selected lines and the control lines, while under directional selection there is at
least a theoretical expectation that genetic variances and covariances would change between
selected and unselected populations \citep[e.g.][]{falconer1996introduction}, and work on the
$31$st generation of mice from this selection experiment has demonstrated some changes in the
genetic variances of wheel running over the first $6$ days of wheel running
\citep[][]{careau2015evolution}. Second, the results suggest that line $1$ randomly differs from
one other control line and one other selected line.  Such random differences in biological
populations can be caused by genetic drift occurring during the selection experiment or by founder
effects when the original base population was randomly subdivided into eight lines. Indeed, the
trait of selection itself (wheel running on days 5 and 6 of a 6 day exposure) and underlying
physiological traits (e.g., basal metabolic rate) demonstrate the effects of drift and/or founder
effects \citep{swallow1998artificial,kane2008basal}. The results presented herein are suggestive of
similar processes influencing the phenotypic covariance structure of wheel running across age,
which presents interesting possibilities of additional biological experiments.

%The result of the test is somehow surprising because there should not be any difference between the
%first line and the other control lines. On the other hand, there is no evidence of a difference
%between the other control lines and the selected lines, which could have been expected. This
%suggests further investigations in the experimental conditions of the first line and some caution
%to draw conclusions when it is included in the analysis.

\section{Conclusions and further developments}
We extended the application of hypothesis tests that take into account the geometry of the space of
covariance operators to the case of multiple groups, using a permutation approach. In particular,
synchronized permutations allow us to make inference also on the pairwise comparison between
groups while controlling the family-wise error rate. We illustrate via simulation studies that the
proposed test has the correct effect size and indeed the square root distance and the Procrustes
distance lead to higher empirical power in the multiple groups comparison as well. While we have
shown that the method can be applied in the case of a missing observation, a more general treatment
of the case of unbalanced design and missing data is scope for future works.

We have also shown that the empirical power for the global test is comparable to those obtained
using bootstrap approximation in the Gaussian case and slightly better in the non Gaussian case. It
is worth to notice that, while simulation results shows the bootstrap approach to be promising as
well for the global test, its property has not yet rigorously studied for test statistics based on
metric different from the Hilbert-Schmidt distance and this is an interesting direction for future
research.

The application of the procedure to the mice voluntary wheel running activity curves shows that,
while a difference between covariance operators is indeed present, this is not caused by selection
itself.  Instead it would appear that random biological processes such as genetic drift or founder
effects are influencing the covariance operators of the phenotypic curves, suggesting further
investigation of this trait and demonstrating the importance of random processed during evolution.
\bibliographystyle{apalike}
\bibliography{../bibliography}

\end{document}